\documentclass[lettersize,journal]{IEEEtran}
\usepackage{amssymb, booktabs, multirow, amsmath}
\usepackage{mathtools}
\usepackage{algorithmic}
\usepackage{algorithm}
\usepackage{array}
\usepackage[caption=false,font=normalsize,labelfont=sf,textfont=sf]{subfig}
\usepackage{textcomp}
\usepackage{stfloats}
\usepackage{url}
\usepackage{verbatim}
\usepackage{graphicx}
\usepackage[noadjust]{cite}
\usepackage{balance}

\usepackage{caption}

\newcommand{\bmX}{\mathcal{X}}
\newcommand{\bmY}{\mathcal{Y}}
\newcommand{\bmZ}{\mathcal{Z}}
\newcommand{\bmS}{\mathcal{S}}

\newcommand{\bmI}{\mathcal I}
\newcommand{\bmL}{\mathcal L}

\newcommand{\reals}{\mathbb{R}}
\newcommand{\Va}{{\mathbf{X}}}
\newcommand{\Ha}{{\mathbf{H}}}

\newcommand{\Wa}{{\mathbf{W}}}

\newcommand{\xtime}{x}
\newcommand{\phase}{\mathbf{P}}

\usepackage{xcolor}

\begin{document}
\title{Tackling Interpretability in Audio Classification Networks with Non-negative Matrix Factorization}
\author{Jayneel Parekh, Sanjeel Parekh, Pavlo Mozharovskyi, Ga\"{e}l Richard, Florence d'Alch\'e-Buc 
\thanks{Manuscript received April, 10 2023; Jayneel Parekh, Sanjeel Parekh, Pavlo Mozharovskyi, Ga\"{e}l Richard, Florence d'Alch\'e-Buc are with the Laboratoire de Traitement et Communication de l’Information (LTCI), T\'el\'ecom Paris, Institut Polytechnique de Paris, 91120 Palaiseau, France. (e-mail: \{jayneel.parekh,pavlo.mozharovskyi,florence.dalche,gael.richard\}@telecom-paris.fr, sanjeelparekh@gmail.com)}}

\markboth{Under submission}%
{Interpretability and Audio Classification Networks with NMF}

\maketitle

\begin{abstract}
  This paper tackles two major problem settings for interpretability of audio processing networks, \textit{post-hoc} and \textit{by-design} interpretation. For post-hoc interpretation, we aim to interpret decisions of a network in terms of high-level audio objects that are also listenable for the end-user. This is extended to present an inherently interpretable model with high performance. To this end, we propose a novel interpreter design that incorporates non-negative matrix factorization (NMF). In particular, an interpreter is trained to generate a regularized intermediate embedding from hidden layers of a target network, learnt as time-activations of a pre-learnt NMF dictionary. Our methodology allows us to generate intuitive audio-based interpretations that explicitly enhance parts of the input signal most relevant for a network's decision. We demonstrate our method's applicability on a variety of classification tasks, including multi-label data for real-world audio and music.
\end{abstract}








%


\section{Introduction}



Deep learning models, while state-of-the-art for several tasks in domains such as computer vision, natural language processing and audio, are  typically not interpretable. Their increasing use, in decision-critical domains, in every-day applications, raises the problem of  interpreting their decisions. This issue has been grappled with in two different ways by the research community. The first one relies on the availability of a predictive model trained and optimized for performance but not for interpretability and aims to devise an additional approach to interpret the given model. This problem setting is usually referred to \textit{post-hoc interpretation}.
The other setting aims to build an interpretable predictive model from the data. The challenge here is to demonstrate high classification performance while maintaining interpretability in the same model. This setting is often referred to as \textit{by-design interpretation} problem. Real-world scenarios of utilizing interpretability of networks can occur under variety of constraints and demands regarding deployment, level of interpretability and performance. Thus, from a practical standpoint, both problem settings hold independent value.


In this paper, our aim is to address both problems for audio classification networks while proposing a system more suited for understanding interpretations for audio modality than other common methods in literature. 

An ideal interpreter is supposed to offer insights about a model's decision in an understandable fashion to humans. In the case of audio classification, there are certain desirable traits for an interpreter that effectively help to fulfil this purpose. Firstly, we advocate that the interpretations should be generated in terms of high-level audio objects. Even more importantly, the interpretation should be listenable for an end-user. The rationale behind posing these traits as desirable is as follows: Audio scenes are often composed of multiple high-level audio objects \cite{bregman1994auditory}. 
Moreover, understanding events/scenes through the notion of audio objects also aligns with cognitive development in human and animals \cite{griffiths2004auditory, dyson2004representation}. Listenability is essential since it is significantly more intuitive and easier to listen to an interpretation rather than visualizing it in its time-frequency representation (eg. spectrogram). Usefulness for both the traits can be reinforced through an example. Imagine an audio-based surveillance system for a house raising an alarm for break-in. An interpreter can be expected to be able to localize the event among a host of concurrent events that triggered the alarm. If for example 'glass-breaking' is the triggering event that the interpreter recognizes in the input, a human would find it easier to understand, if they can hear the interpretation rather than visualize it on a spectrogram. 

To this end, we propose an interpreter that relies on processing selected hidden representations of the classifier by a neural network to extract an intermediate embedding. This intermediate encoding is regularized in multiple ways, the two essential ones being: (i) Mimicking the classifier output to be able to interpret its decisions, and (ii) Reconstruct the input through the help of a dictionary of spectral patterns. The latter loss and its design is strongly inspired by the structure in Non-negative Matrix Factorization (NMF, \cite{lee2001nmf}), known to provide part-based decompositions. The loss is crucial in imposing a highly understandable meaning of ``time activations" on the intermediate embedding. This decomposition structure also allows the interpreter to benefit from filtering information from the input. It’s worth emphasizing that audio interpretability is not the same as classical tasks of separation or denoising. These tasks involve recovering complete object of interest in the output audio. On the other hand, a classifier network might focus more on salient regions. When interpreting its decision and making it listenable we expect to uncover such regions and not necessarily the complete object
of interest.    

This paper is an extension of our work on post-hoc interpretability \cite{l2i}. It differs in two significant ways: (a) In addition to post-hoc, we present a by-design interpretable network that has state-of-the-art classification performance among interpretable models for audio. To do so, we modify the training procedure to allow the hidden layers of the classifier to be fine-tuned, and apply a classification loss at its output. The interpreter's architecture and objectives remain the same for both problems. (b) We demonstrate the utility of the proposed methods on music data with results on largest publicly available polyphonic music instrument recognition dataset, OpenMIC-2018 \cite{openmic}.






In summary, we make the following contributions:    

\begin{itemize}

    \item We build a holistic approach that generates listenable concept-based interpretations to tackle post-hoc and by-design interpretability for audio classification networks.

    \item We present an original formulation that constrains the interpreter encoding through two loss functions, one for input reconstruction through NMF dictionary and the other for fidelity to the network's decision. From a learning perspective, we show a new way to link NMF with deep neural networks, especially for generating interpretations.
    \item We extensively evaluate on three popular audio event analysis benchmarks, tackling both multi--class and multi--label classification tasks. The dataset for the latter is very challenging due to its collection in noisy real--world settings. Our method's design allows us to simulate feature removal and perform \textit{faithfulness} evaluation.
\end{itemize}




    
    
    
    
    

\section{Related Works}

We organize the discussion about literature in three parts. We first cover interpretability works in a broad sense, followed by audio specific methods. We conclude the section by discussion on applications of NMF to audio signal processing. 

\subsection{General Interpretability literature} 

{\bf Feature attribution}: The vast majority of interpretability literature is covered under feature attribution methods. They are a class of methods which offer interpretations through input feature importance or selection. In case of post-hoc interpretation, this includes perturbation based approaches \cite{lime, shap, muse}, as well as saliency map based approaches \cite{guidedbackprop, smoothgrad, iba, lrp, gradcam}. Perturbation based approaches rely on observing model output on many locally perturbed versions of the input to determine importance of each individual feature for the decision. Part of the research challenge for these methods is to define ``meaningful" perturbations \cite{socrat}. Saliency map based approaches typically generate interpretations through modified gradient backpropagation \cite{guidedbackprop, lrp}, but some also utilize upsampled versions of intermediate activation maps \cite{gradcam, iba}. This form of interpretation is also common for by-design interpretable models. Common ways of training such models involve modifying the architecture, loss function (or both) to incorporate interpretability in the model.   

{\bf Interpretation beyond attribution}: Feature attribution methods have been under scrutiny for their robustness and faithfulness \cite{kindermans2017unreliability, robust-saliency}. This has overlapped with increasing amount of research for both post-hoc and by-design interpretation models to develop systems that offer interpretations through different means. Prior research has now proposed systems that provide interpretations via logical rules \cite{rrn}, counterfactuals \cite{steex} and even natural language \cite{hendricks2016}. Each of them can potentially be a more suited choice for certain problem domains and use-cases. Our approach uses high-level audio objects for interpretation. This aspect renders it closer to prototype and concept-based approaches. Prototype-based approaches \cite{protodnn, chen2019looks} tackle by-design interpretability by learning a predefined number of embeddings which are encouraged to represent singular training datapoints. The final classification decision is made based on similarity of given test sample to all the prototypes. Concept-based approaches aim to represent high-level concepts explicitly and subsequently offer interpretation in terms of them. An interesting approach is that of FLINT \cite{flint} with whom we share the idea of utilizing the hidden layers and loss functions to encourage interpretability. However we crucially differ from FLINT and other related approaches in concept representation and their applicability for audio interpretations. FLINT represents concepts by a dictionary of attribute functions over input space. The learnt concepts are not obviously comprehensible to a user, requiring a separate visualization pipeline to get insights. Approaches based on TCAV \cite{kim17tcav}, such as ACE \cite{ace}, ConceptSHAP \cite{conceptshap}, define concept using a set of images and learn a representation for it in terms of hidden layers of the network, termed as concept activation vector (CAV). These designs for concepts are not related to our NMF-inspired dictionary representation. Importantly, none of the above mentioned approaches can generate listenable interpretations which is key for understandability of audio processing networks.   

\subsection{Audio Interpretability}
\label{related:audio_intp}

Compared to literature in other domains like text or images, work on interpretability with audio signals is considerably sparse. Usefulness of saliency-map based methods for audio interpretability has been illustrated by prior works. Becker et al. \cite{aud-lrp} demonstrate the use of popular Layerwise Relevance Propagation (LRP) \cite{lrp} algorithm for audio digit recognition task. Won et al. \cite{att-int} highlight the use of attention for music tagging task. Muckenhirn et al. \cite{muckenhirn2019relevance} use GuidedBackprop \cite{guidedbackprop} to analyze CNNs operating on 1D waveform. However, these methods do not address the issue of listenability of interpretations and moreover are limited to post-hoc interpretation. A few works based on the LIME algorithm \cite{lime} have attempted to address this issue. SLIME \cite{slime-1, slime-2} proposed to segment the input along time or frequency. The input is perturbed by switching "on/off" the individual segments. AudioLIME \cite{aud-lime-1, aud-lime-2} proposed to separate the input using predefined sources to create the simplified representation. AudioLIME arguably generates more meaningful interpretations than SLIME as it relies on audio objects readily listenable for end-users. However, it suffers from limited applicability, requiring existence of known and meaningful predefined sources that compose the input audio. More recently, \cite{music_concepts} extended the idea of TCAV to represent concepts in music data. The supervised approach requires the overhead of human annotation of concepts, whereas the unsupervised approached based on non-negative tensor decomposition faces the challenge of meaningful learning of concepts. APNet \cite{apnet} proposes an interpretable system by-design that extends prototypical networks \cite{chen2019looks, protodnn} for audio input by defining a more suitable distance measure for audio prototypes. An advantage of our approach over all the previous methods is its ability to address both post-hoc and by-design problem settings. Another unique feature of the approach is its means of generating interpretations, relying on a dictionary of NMF components for the same. From the point of view of interpretability, this is a novel strategy to gain insights about a model.

\subsection{NMF applications for audio} 
NMF is a data decomposition technique popularized by Lee and Seung \cite{lee2001nmf} as a method to learn ``parts of an object''. It has since been used widely within the audio community to tackle source separation \cite{NMF_smaragdis1}, denoising \cite{NMF_smaragdis2}, inpainting \cite{le2011computational} and transcription \cite{NMF_transcription1, NMF_transcription2}. Typically, a nonnegative time-frequency audio representation is decomposed into two nonnegative factors namely, spectral patterns or dictionary matrix and their time activations. Its traditional usage as a supervised dictionary or feature learning method involves learning class-wise dictionaries over training data \cite{fevotte2018single}. Time activations, the so-called features, can then be generated for any input by projecting it onto the learnt dictionaries. These features can subsequently be used for downstream tasks such as classification. In this section we focus on prior works that combine NMF with deep architectures.

Bisot et al. \cite{bisot2017feature} couple NMF-based features with neural networks to boost performance of acoustic scene classification. NMF has also been successfully employed with audio--visual deep learning models for separation \cite{gao2018learning} and classification \cite{parekh2019identify}.

Iterations of NMF optimization algorithms can be unfolded as novel deep neural networks. This observation has led to development of ``Deep NMF'' methods. In particular, Le Roux et al. \cite{le2015deep} unfold the multiplicative updates of NMF parameters into a deep network for speech separation. Wisdom et al. \cite{wisdom2017deep} apply this strategy to iterative soft thesholding algorithm to propose deep recurrent NMF.

While these works share with us the high-level idea of combining neural networks and NMF, there is no overlap between our goals and methodologies. Unlike aforementioned studies, we wish to investigate a classifier's decision using NMF as a regularizer. Furthermore, to our best knowledge, attempting to regress temporal activations of a fixed NMF dictionary by accessing intermediate layers of an audio classification network is novel even within the NMF literature.



\section{System Design}


\begin{figure*}[t]
\centering
\includegraphics[width=0.6\textwidth]{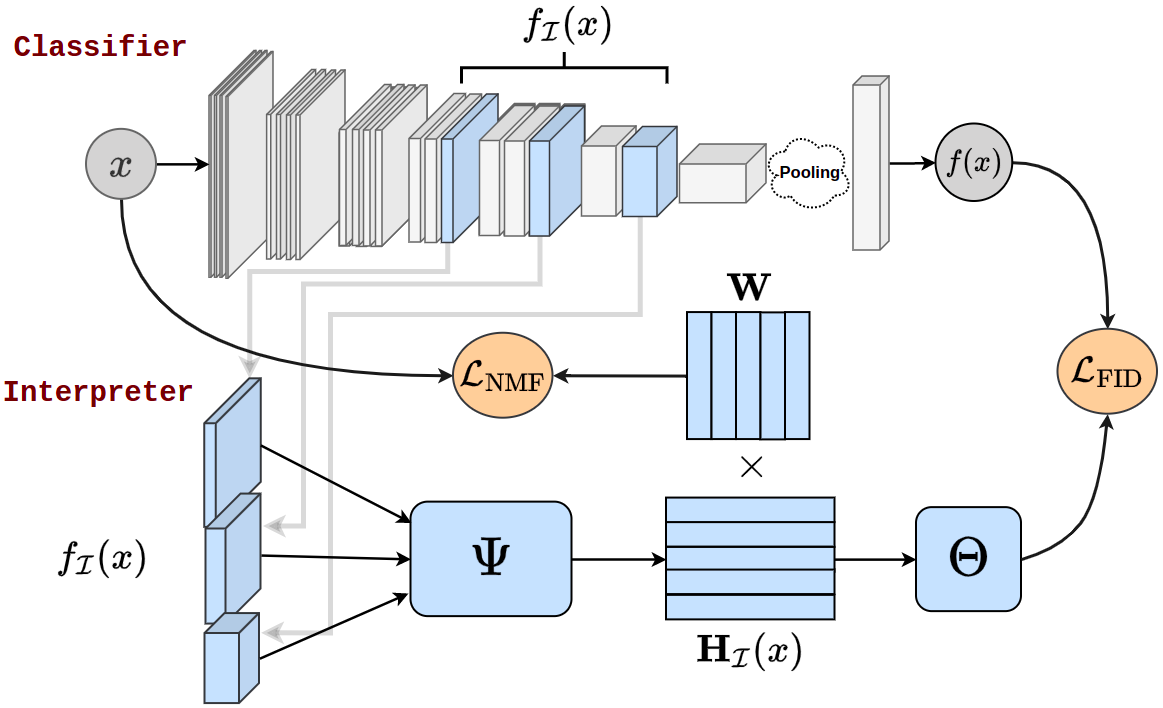}
\caption{\textbf{System overview}: The core design common to both post-hoc and by-design interpretation. The interpreter (indicated in blue) accesses hidden layer outputs of the classifier. These are used to predict an intermediate encoding. Through regularization terms, we encourage this encoding to both mimic the classifier's output and also serve as the time activations of a pre-learnt NMF dictionary. In post-hoc interpretation, the classifier is pre-trained and fixed, and only the interpreter is trained. For by-design interpretation we train both jointly and make final predictions using output of interpreter. 
}
\label{fig_sys}
\end{figure*}

We organize this section as follows: We start with a brief note on notation used throughout the paper. We describe the setup of our framework to address post-hoc interpretation in section \ref{approach:post-hoc}. This is extended to address by-design interpretation in section \ref{approach:by-design}. We expound on the specific architectural details common to both problem settings in section \ref{approach:fillgaps} and conclude the section by detailing how we generate interpretations with our design in section \ref{interpretation_generation}.  

\subsection{Data Notation}
\noindent We denote a training dataset by $\bmS := (\xtime, y)_{i=1}^N$, where $\xtime \in \bmX$ is the time domain audio signal and $y \in \bmY$, a label vector. The label vector could be a one-hot or binary encoding depending upon a multi-class or multi-label dataset, respectively. 
For listenable interpretations through NMF, we favor a representation of $x$ that can be easily inverted back to the time-domain and use a log--magnitude spectrogram $\mathbf{X} \in \mathbb{R}^{F \times T}$ that is computed  by applying an element-wise transformation $x_0 \rightarrow \log(1 + x_0)$ on the magnitude spectrogram with $F$ frequency bins and $T$ time frames. This is preferred over using magnitude spectrograms as it corresponds more closely to human perception of sound intensity \cite{goldstein1967auditory}. A deep neural network classifier for post-hoc interpretations is denoted as $f: \bmX \rightarrow \bmY$.  

\subsection{Post-hoc Interpretation}
\label{approach:post-hoc}

When addressing the problem of post-hoc interpretation, the classifier $f$ will be pre-trained and then fixed throughout. We describe now the components of the interpreter and what are its inputs and outputs. 


\noindent \textbf{Overview} The system design is illustrated in figure \ref{fig_sys}. The interpreter is designed to have access to hidden representations of the classifier and is tasked to produce an intermediate embedding through the function $\Psi$. This embedding is placed under certain constraints via function $\Theta$ and a pre-learnt dictionary of NMF components $\Wa$. These constraints impose a highly meaningful structure on the embedding and help in interpreting the decision $f(x)$. We discuss the constrains, which form the core of our approach, in this subsection. The precise architectures of $\Psi$ and $\Theta$ and optimization problem used to pre-learn $\Wa$ are covered later in Sec. \ref{approach:fillgaps}.


Specifically, hidden layer outputs of the classifier $f$, taken as input by the interpreter, are denoted as $f_{\bmI}(x) \in \bmZ$. They are processed through the function $\Psi: \bmZ \rightarrow \mathbb{R}^{K \times T}_{+}$, modelled as a neural network. This produces an intermediate encoding. For simplicity, we will denote this encoding generated from hidden layers as $\Ha_{\bmI}(x) = \Psi \circ f_{\bmI}(x)$, a function over input $x$. The constraints on this encoding, implemented as loss functions are as follows:

\textbf{Loss 1 (Fidelity loss):} To be able to identify the relevant signal for interpretation, we constrain $\Ha_{\bmI}(x)$ to approximate classifiers output probabilities $f(x)$ through the function $\Theta: \mathbb{R}^{K \times T}_{+} \rightarrow \bmY$. The term $\Theta(\Ha_{\bmI}(x))$ is also referred to as interpreter's output. We implement this constraint as a loss function by minimizing the generalized cross-entropy loss between $\Theta(\Ha_{\bmI}(x))$ and $f(x)$. We refer to it as the fidelity loss $\bmL_{\textrm{FID}}$. Denoting the parameters of $\Psi, \Theta$ as $V_{\Psi}, V_{\Theta}$, for multi-class classification the loss can be written as,
\begin{equation}
    \bmL_{\text{FID}}(x, V_\Psi, V_\Theta) = -f(x)^\intercal\log(\Theta(\Ha_{\bmI}(x)))
\end{equation}
On the other hand, for multi-label classification this loss reads,
\begin{equation}
    \begin{aligned}
    \bmL_{\text{FID}}(x, V_\Psi, V_\Theta) = &-\sum f(x) \odot \log(\Theta(\Ha_{\bmI}(x)))\\
    & + (1-f(x)) \odot  \log(1-\Theta(\Ha_{\bmI}(x))).
    \end{aligned}
\end{equation}
Here $\odot$ denotes element-wise multiplication.

\textbf{Loss 2 (Reconstruction loss):} We additionally constrain $\Ha_{\bmI}(x)$ to be able to reconstruct the input audio using pre-learnt dictionary $\Wa \in \reals^{F \times K}_{+}$. This constraint asks to decompose input log-magnitude spectrogram as $\Va \approx \Wa\Ha_{\bmI}(x)$, that is, a product of two non-negative matrices. This loss is based on popular non-negative matrix factorization. Crucially, this allows us to consider $\Ha_{\bmI}(x)$ as a time activation matrix for $\Wa$. We refer to this as the reconstruction loss, denoted as $\bmL_{\textrm{NMF}}$.
\begin{equation}
    \bmL_{\text{NMF}}(x, V_{\Psi}) = \|\mathbf{X} - \Wa\Ha_{\bmI}(x)\|^2_2.
\end{equation}

\textbf{Loss 3 (Sparsity loss):} In addition to $\bmL_{\text{FID}}$ and $\bmL_{\text{NMF}}$, we impose $\ell_1$ regularization on $\Ha_{\bmI}(x)$ to encourage well-behavedness, especially for large dictionary sizes \cite{le2015sparse}.

\noindent \textbf{Training optimization}.~  The complete loss function over our training dataset $\bmS$ can thus be given as:
\begin{equation}
    \bmL(V_\Psi, V_\Theta) = \sum_{x \in \bmS} \bmL_{\text{FID}}(x, V_\Psi, V_\Theta) + \alpha\bmL_{\text{NMF}}(x, V_{\Psi}) + \beta||\Ha_{\bmI}(x)||_1
    \label{all_loss}
\end{equation}
where $\alpha, \beta \geq 0$ are loss hyperparameters. All the parameters of the system are constituted in the functions $\Psi, \Theta$ and dictionary $\Wa$. Since $\Wa$ is pre-learnt and fixed, the training loss $\bmL$ is optimized only w.r.t  $V_\Psi, V_\Theta$. As a reminder, when training the interpreter for post-hoc analysis, the classifier network is kept fixed. The final optimization problem addressed for post-hoc interpretation writes as follows:
\begin{equation}
    \hat{\Psi}, \hat{\Theta} = \arg \min_{\Psi, \Theta} \bmL(V_{\Psi}, V_{\Theta})
\end{equation}
\subsection{By-design Interpretation}
\label{approach:by-design}

Interestingly, the same framework can also be utilized to train an inherently interpretable model. As a first step, we propose the following function to be used for making final predictions
\begin{equation*}
    g: \bmX \rightarrow \bmY, g(x) = \Theta \circ \Psi \circ f_{\bmI}(x)
\end{equation*} 
which is a mixture of interpreter and classifier layers. One might be tempted to employ the same training mechanism for by-design problem as done for post-hoc interpretation. However, there is a difference in the problem setting we need to adapt for. Namely, the classifier layers are not trained for prediction as before. This implies that we cannot simply aim to generate meaningful representations from it as is. 

To remedy the difficulty, we modify the training in two different ways: (i) Layers of $f$ are now modified by backpropagating all interpreter losses to them, and thus are now jointly trained with the interpreter. (ii) We modify the loss function to include an additional prediction loss on the output $f(x)$ to train all the layers in $f$. The training loss function and optimization problem write as following:

\begin{equation*}
    \begin{aligned}
    & \bmL_{f}(x, V_f) = -y^\intercal\log(f(x))\\
    & \bmL_{\text{NMF}}(x, V_\Psi, V_f) = \|\mathbf{X} - \Wa\Ha_{\bmI}(x)\|^2_2\\
    & \bmL_{\text{FID}}(x, V_\Psi, V_\Theta, V_f) = -f(x)^\intercal\log(\Theta(\Ha_{\bmI}(x)))
    \end{aligned}
\end{equation*}
\begin{equation*}
    \begin{aligned}
    \bmL(V_\Psi, V_\Theta, V_f) = &\sum_{x \in \bmS} \bmL_f(x, V_f) + \gamma\bmL_{\text{FID}}(x, V_\Psi, V_\Theta, V_f)\\
    & + \alpha\bmL_{\text{NMF}}(x, V_{\Psi}, V_f) + \beta||\Ha_{\bmI}(x)||_1    
    \end{aligned}
\end{equation*}
\begin{equation*}
    \hat{\Psi}, \hat{\Theta}, \hat{f} = \arg \min_{\Psi, \Theta, f} \bmL(V_{\Psi}, V_{\Theta}, V_{f})   
\end{equation*}

A reader might question the need for applying a prediction loss at output of $f$ when the function $g$ described above is proposed to make final predictions. This is indeed a reasonable variant of our current choice and we resolve this issue by comparing the performance of both systems in experiments in section \ref{experiments:by-design} 







\subsection{Filling the gaps}
\label{approach:fillgaps}

It should be noted that the network architectures and other implementation details remain the same in both problem settings. We now cover the remaining architectural details of $\Psi, \Theta$ and the algorithm for pre-learning $\Wa$.

\noindent \textbf{Design of $\Psi$}.~ The network $\Psi$ is tasked with producing the encoding $\Ha_{\bmI}(x) \in \reals_{+}^{K \times T}$ from the set of convolutional feature maps of the classifier, given by $f_{\bmI}(x)$. These feature maps potentially originate from different layers and thus can be of different resolutions. To perform joint processing on them, each one is first appropriately transformed to ensure same width and height dimensions. The subsequent layers process these maps through some convolutional (with ReLU activation) and resampling layers. However, this composition is based on certain important aspects. Firstly, audio feature maps of CNNs with spectrogram-like inputs contain the notion of time and frequency along the width and height dimensions. Secondly, our goal with this network is to process a 3D representation of feature patterns across time and frequency, and convert it to a 2D intermediate encoding that can serve as time activation matrix of size $K \times T$. To achieve this, the subsequent convolutional layers continuously decrease resolution on the frequency axis and increase resolution the time axis to $T$ frames. Furthermore, the input axis for number of feature maps corresponds to the axis of number of components $K$ in output of $\Psi$, equal to the number of components in dictionary $\Wa$.



\noindent \textbf{Design of $\Theta$}.~ The goal of this network is to mimic the output $f(x)$ by processing $\Ha_{\bmI}(x)$ This directly helps in shaping $\Ha_{\bmI}(x)$ to interpret $f(x)$. An important consideration for designing $\Theta$ was to keep its operations on $\Ha_{\bmI}(x)$ interpretable. This helps during the interpretation phase in easily quantifying how different parts of $\Ha_{\bmI}(x)$ influence the interpreters output. It is thus composed of two parts. The first part pools activations $\Ha_{\bmI}(x)$ across time. This pooling can be implemented in multiple ways, for eg. max or average pooling. However, we opt for an intermediate style of attention--based pooling \cite{mil-att}, \textit{i.e.}, $\mathbf{z} = \sum_{t=1}^{T} \Ha_{\bmI}(x)\mathbf{a}$, where $\mathbf{a} \in \reals^{T}$ are the attention weights and $\mathbf{z} \in \reals^{K}$ is the pooled vector. The pooled representation vector is passed through a linear layer. This is followed by an appropriate activation function to convert its output to probabilities, that is, softmax for multi-class classification and sigmo\"{i}d for multi-label classification.

\noindent \textbf{Pre-learning $\Wa$}.~ The non-negative matrix $\Wa$ forms an integral part of the interpreter design. It is pre-learnt from the input data, and essential in formulating the reconstruction loss $\bmL_{\text{NMF}}$. We employ Sparse-NMF \cite{le2015sparse} for the pre-learning. The following optimization problem is solved through multiplicative updates to pre-learn $\Wa$:
\begin{equation}
	\begin{aligned}
		&\text{min} & & D(\Va_{\text{train}}|\Wa\Ha) + \mu\|\Ha\|_1\\
			& \text{subject to} & & \Wa \geq 0, \Ha \geq 0,\\
			& & & \|\mathbf{w}_k\|=1,~\forall k.
    \end{aligned}
    \label{sparse-nmf}
\end{equation}
where $\Va_{\text{train}}$ is a subset of the training data $\bmS$. Note that its construction is dataset dependent and will be covered in experiments. Here $D(.|.)$ is a divergence cost function. In practice, euclidean distance is used. Training audio files are converted into log-magnitude spectrogram space for factorization.

\subsection{Generating Interpretations}
\label{interpretation_generation}

Having described the goals and details of all components of our framework, we finally discuss how the interpretations are generated. To generate audio that interprets the classifier's decision for a sample $x$ and a predicted class $c$, we follow a two-step procedure: The first step consists of identifying the components which are considered ``important" for the prediction. This is determined by estimating their relevance using the pooled time activations in $\Theta$ and the weights for linear layer. Precisely, given a sample $x$, the pooled activations are computed as $\mathbf{z} = \Ha_{\bmI}(x)\mathbf{a}$. Denoting the weights for class $c$ in the linear layer as $\theta_c^w$, the relevance of component $k$ is estimated as $r_{k, c, x} = \frac{(\mathbf{z}_k\theta_{c, k}^w)}{\max_l |\mathbf{z}_l\theta_{c, l}^w|}$. This is essentially the normalized contribution of component $k$ in the output logit for class $c$. To select the ``important" components, we simply threshold the relevance via a parameter $\tau \in (0,1)$ as, $L_{c, x} = \{ k: r_{k, c, x} > \tau \}$.

    
    
    The second step consists of estimating a time domain signal for each relevant component $k \in L_{c, x}$ and also for set $L_{c, x}$ as a whole. In this paper, we refer to the latter as the generated interpretation audio, $x_{\text{int}}$. For certain classes, it may also be meaningful to listen to each individual component, $x_k$. As discussed earlier under NMF basics, estimating time domain signals from spectral patterns and their activations typically involves a soft--masking and inverse STFT procedure. We detail this step with appropriate equations in Algorithm \ref{alg:gen}.
    



\begin{algorithm}
\linespread{1.2}\selectfont
\caption{Audio interpretation generation}\label{alg:gen}
\begin{algorithmic}[1]
\STATE \textbf{Input:} log-magnitude spectrogram $\mathbf{X}$, input phase $\phase_x$ components $\Wa = \{ \mathbf{w}_1, \ldots, \mathbf{w}_K \}$, time activations $\Ha_{\bmI}(x) = [\mathbf{h}^{\bmI}_1(x), \ldots, \mathbf{h}^{\bmI}_K(x)]^\intercal$, set of selected components $L_{c, x}=\{ k_1, \ldots, k_B \} $.

\FORALL {$k \in L_{c, x}$}
\STATE $\mathbf{X}_k \leftarrow \frac{\mathbf{w}_k\mathbf{h}^{\bmI}_k(x)^\intercal}{\sum_{l=1}^{K}\mathbf{w}_l\mathbf{h}^{\bmI}_l(x)^\intercal} \odot \mathbf{X}$ \hfill \algorithmiccomment{// Soft masking}
\STATE $x_k = \textrm{INV}(\mathbf{X}_{k}, \phase_x)$     \hfill \algorithmiccomment{// Inverse STFT}
\ENDFOR
\STATE $\mathbf{X}_{\text{int}} \leftarrow \sum_{k \in L_{c, x}} \mathbf{X}_k$
\STATE $x_{\text{int}} = \textrm{INV}(\mathbf{X}_{\text{int}}, \phase_x)$
\STATE \textbf{Output: }$\{x_{k_1}, \ldots, x_{k_B}\}$, $x_{\text{int}}$
\end{algorithmic}
\end{algorithm}

\section{Experimental design}

Most of the experimental settings remain the same for post-hoc and by-design interpretations since the underlying architecture and the loss functions directly affecting interpreter are identical. Thus, the datasets, audio representation used by the network and the learnt dictionaries remain unchanged. However, there are some differences in training and evaluation that will be discussed explicitly. We start by covering the above details in section \ref{design:datasets}-\ref{design:impl}. We discuss the interpretation evaluation strategies relevant for both problems in section \ref{metrics}, including all the systems evaluated. 

\subsection{Datasets}
\label{design:datasets}
We experiment with three datasets covering different types of learning tasks, source data etc. We discuss each of them in greater detail below. 
\subsubsection{ESC50} ESC-50 \cite{esc50} is a popular benchmark for environmental sound classification task. It is a multi-class dataset that contains 2000 audio recordings of 50 different environmental sounds. The classes are broadly arranged in five categories namely, animals, natural soundscapes/water sounds, human/non-speech sounds, interior/domestic sounds, exterior/urban noises. Each clip is five-seconds long and extracted from publicly available recordings on the \texttt{freesound.org} project. The dataset is prearranged into 5 folds.

\subsubsection{SONYC-UST} The DCASE task used a very challenging real-world dataset called Sounds of New York City-Urban Sound Tagging (SONYC-UST) \cite{cartwright2019sonyc}. It contains audio collected from multiple sensors placed in the New York City to monitor noise pollution. It consists of eight coarse-level and 20 fine-level labels. We opt for the coarse-level labeling task that involves multi-label classification into: `engine', `machinery-impact', `non-machinery-impact', `powered-saw', `alert-signals', `music', `human-voice', `dog'. 
This task is highly challenging for several reasons: (i) since it is real-world audio, the samples contain a very high level of background noise, (ii) the audio sources corresponding to the classes are often weak in intensity, as they are not necessarily close to the sensors, (iii) some classes may also be highly localized in time and more challenging to detect, (iv) lastly, noisy audio also makes it difficult to annotate, leading to labeling noise. This is especially true for training data labeled by volunteers.

\subsubsection{OpenMIC-2018} The OpenMIC-2018 dataset \cite{openmic} is composed of 20000 polyphonic audio recordings annotated with weak labels from among 20 instrument classes. The dataset was created by querying the content available on Free Music Archive under the Creatives Commons license with AudioSet concept ontology and using a multi-instrument estimator model trained on AudioSet data to suggest candidates for annotation. Each recording/clip is 10 seconds long. A single sample generally consists of weak labels of only a small subset of classes. Each instrument class has at least 500 positive and 1500 total annotated samples. Compared to SONYC-UST, the number of positive samples intra class and inter class are considerably more balanced. It is currently the only large publicly available dataset with multi-label annotation for polyphonic audio.

\subsection{Implementation details}
\label{design:impl}
\subsubsection{Classification network} We interpret a VGG-style convolutional neural network proposed by Kumar et al. \cite{kumar_wft}. This network was chosen due to its popularity and applicability for various audio scene and event classification tasks. It can process variable length audio and has been pretrained on AudioSet \cite{audioset}, a large-scale weakly labeled dataset for sound events. It takes as input a log-mel spectrogram. The architecture broadly consists of six convolutional blocks (B1--B6) followed by a convolutional layer with pooling for final prediction. Most convolutional blocks consist of two sets of conv2D + batch norm + ReLU layers followed by a max pooling layer. We fine-tune this network on each dataset separately before training our system for any post-hoc interpretations. For ESC-50, we modify only fully-connected layers after the convolutional blocks while for SONYC-UST and OpenMIC-2018, we modify all the layers during fine-tuning. 

\noindent \textbf{Classifier performance}.~ On ESC-50, the classifier is evaluated using 5-fold cross-validation. It achieves an accuracy of $82.5 \pm 1.9$\% over the 5 folds, higher than the average human accuracy of 81.3\%. SONYC-UST is an unbalanced multi-label dataset. The evaluation is done using AUPRC based metrics. Our fine-tuned classifier achieves a macro-AUPRC (official metric for DCASE 2020 challenge) of $0.601$. This is higher than the DCASE baseline performance of 0.510 and comparable to the best performing system macro-AUPRC of 0.649 \cite{Arnault2020}. Note that it is obtained without use of data augmentation or additional strategies to improve performance. OpenMIC-2018 is a relatively balanced multi-label dataset. To evaluate our trained classifier, we use the weighted average F1-score metric, proposed in the original paper. The metric computes for each class a weighted average of F1-scores over the positive and negative samples. The final score is the average over 20 classes. Our classifier achieves final score of 0.83, better than the VGGish based baseline score of 0.78 and competitive with other recent models. These details are tabulated in Tab. \ref{benchmark_f}. As noted earlier, the pre-training is only executed for post-hoc interpretations.

\begin{table}
    \setlength{\tabcolsep}{3pt}
	\centering
    \resizebox{0.49\textwidth}{!}{%
	\begin{tabular}{l c c c} 
		\toprule
		& \multicolumn{1}{c}{ESC-50 (in \%)} & \multicolumn{1}{c}{SONYC-UST} & \multicolumn{1}{c}{OpenMIC-2018}\\
		\cmidrule[1pt](lr){2-2}
            \cmidrule[1pt](lr){3-3}
            \cmidrule[1pt](lr){4-4}
		System & top-1 & macro-AUPRC & avg-weighted-F1 \\ [0.5ex] 
		\midrule 
            Human accuracy \cite{esc50} & 81.3 & $\times$ & $\times$ \\
            ESC50-CNN baseline \cite{esc50} & 64.5 & $\times$ & $\times$ \\
            Arnault et al. \cite{Arnault2020} & $\times$ & 0.649 & $\times$ \\
            Koutini et al. \cite{bl2_omic} & $\times$ & $\times$ & 0.822 \\
            VGGish \cite{sonyc_ust, openmic} & $\times$ & 0.510 & 0.785 \\
            \midrule
            Current-$f$ & 82.5 & 0.601 & 0.831 \\
 		\bottomrule
	\end{tabular}
     }
	\caption{Benchmarking performance of pre-trained classifier $f$ for post-hoc interpretation.}
    \label{benchmark_f}
\end{table}

\subsubsection{Audio time-frequency representation} For both the tasks, we perform the same audio pre-processing steps. All audio files are sampled at 44.1kHz. STFT is computed with a 1024-pt FFT and 512 sample hop size, which corresponds to about 23ms window size and 11.5ms hopsize. The log-mel spectrogram is extracted using 128 mel-bands.

\subsubsection{Dictionary learning} The matrix on which we apply sparse-NMF to learn $\Wa$, $\Va_{\text{train}}$, is constructed differently for each dataset due to their specific properties. For ESC-50,  $\Va_{\text{train}}$ is constructed by concatenating the log--magnitude spectrograms corresponding to each sample in the training data of the cross-validation fold (1600 samples for each fold). SONYC-UST however, is an imbalanced multilabel dataset with very strong presence of background noise. A procedure to learn components, as for ESC-50, yields many components capturing significant background noise, affecting understandability of interpretations. Hence, we process this dataset differently. We first learn $\Wa_{\text{noise}}$, that is, a set of 10 components to model noise using training samples with no positive label. Then, for each class, we randomly select 700 positively-labeled samples from all training data and learn 10 new components (per class) with $\Wa_{\text{noise}}$ held fixed for noise modeling. All $10 \times 8=80$ components are stacked column-wise to build our dictionary $\mathbf{W}$. While this strategy helps us reduce the number of noise-like components in the final dictionary, it does not completely avoid it. OpenMIC is instead a balanced multilabel dataset for rare noise presence. We simply select random 500 positively labeled samples for each of the 20 classes and learn 15 components. All of them are stacked together to create $\Va_{\text{train}}$.

\subsubsection{Hyperparameters} The hidden layers input to the interpreter module are selected from the convolutional block outputs. As is often the case with CNNs, the latter layers are expected to capture higher-order features. We thus select the last three convolutional block outputs as input to the network $\Psi$. The loss weights and number of components used for post-hoc interpretation are summarized in table \ref{hyperparams:ph}. Ablation studies about all the hyperparameters and justification of their choices will be presented in the next section. The hyperparameters for by-design interpretation are guided by choices in post-hoc interpretation and are tabulated in table \ref{hyperparams:bd}.

\begin{table}
    \setlength{\tabcolsep}{6pt}
	\centering
    \resizebox{0.35\textwidth}{!}{%
	\begin{tabular}{l c c c c} 
		\toprule
		  Dataset & $\alpha$ & $\beta$ & $K$ & \# of epochs \\ [0.5ex] 
		\midrule 
            ESC-50 & 10.0 & 0.8 & 100 & 35 \\
            SONYC-UST & 10.0 & 0.8 & 80 & 21 \\
            OpenMIC-2018 & 5.0 & 0.2 & 300 & 21 \\
            \bottomrule
	\end{tabular}
     }
	\caption{Hyperparameters for all datasets for post-hoc interpretation}
    \label{hyperparams:ph}
\end{table}

\begin{table}
    \setlength{\tabcolsep}{6pt}
	\centering
    \resizebox{0.4\textwidth}{!}{%
	\begin{tabular}{l c c c c c} 
		\toprule
		  Dataset & $\gamma$ & $\alpha$ & $\beta$ & $K$ & \# of epochs \\ [0.5ex] 
		\midrule 
            ESC-50 & 1.0 & 3.0 & 0.2 & 100 & 51 \\
            SONYC-UST & 1.0 & 4.0 & 0.2 & 80 & 21 \\
            OpenMIC-2018 & 1.0 & 3.0 & 0.2 & 300 & 21 \\
            \bottomrule
	\end{tabular}
     }
	\caption{Hyperparameters for all datasets for by-design interpretation}
    \label{hyperparams:bd}
\end{table}

\subsubsection{Optimization}

All the networks are optimized using Adam \cite{adam} with learning rate $2 \times 10^{-4}$.

\subsection{Evaluating Interpretations}
\label{metrics}

Quantifying different aspects of interpretability has been a challenging research question recently. This challenge stems from the inherent subjectivity involved in its definition. Our unique style of ``concept-like" basis for interpretation and global approximation of the base model results in a testing situation to conduct its evaluation, wherein no other method can be directly compared to it. We resolve this hurdle by evaluating different aspects of the interpretation separately. We first discuss quantitative metrics for post-hoc and by-design interpretation along with their goals, followed by discussion on subjective evaluation of interpretations.   

\noindent \textbf{Metrics and baselines (Post-hoc)}.~  The simplest aspect to evaluate is how well does the interpreter agree with the classifier's output. We refer to this metric as the \textit{fidelity} metric. To do so for any given task, we utilize the same metric used to evaluate the classifier performance but instead treat classifiers output as ground truth and evaluate the interpreter's approximation $\Theta(\Ha_{\bmI}(x))$ w.r.t to it. Thus, for multi-class classification, this is done by computing fraction of samples where the class predicted by $f$ is among the top-$k$ classes predicted by the interpreter, referred to as \textit{top-$k$ fidelity}. For multi-label classification tasks with unbalanced number of positive samples of classes, we compute Area Under Precision-Recall Curve (AUPRC) based metrics. In case of balanced classes, we compute F1-score based metrics. We denote our proposed Listen to Interpret (L2I) system, with attention based pooling in $\Theta$ by L2I w/ $\Theta_{\textsc{att}}$. The most suitable baselines to benchmark its fidelity are \textit{post-hoc} methods that approximate the classifier over input space with a single surrogate model. We select two state-of-the-art systems, FLINT \cite{flint} and VIBI \cite{vibi}. A variant of our own proposed method, L2I w/ $\Theta_{\textsc{max}}$, is also evaluated. Herein, attention is replaced with 1D max-pooling operation.


We also conduct a \textit{faithfulness} evaluation for our interpretations. In general for any interpretability method, \textit{faithfulness} tries to assess if the features identified to be of high relevance are \textit{truly} important in classifier's prediction \cite{senn}. Since a ``ground-truth" importance measure for features is rarely available, attribution based methods evaluate faithfulness by performing feature removal (generally by setting feature value to 0) and observing the change in classifier's output \cite{senn}. However, it is hard to conduct such evaluation for non-attribution or concept based interpretation methods on data modalities like image/audio, as simulating feature removal from input is not evident in these cases. 

Interestingly, our interpretation module design allows us to simulate removal of a set of components from the input. Given any sample $x$ with predicted class $c$, we remove the set of relevant components $L_{c, x} = \{k: r_{k, c, x} > \tau \}$ by creating a new time domain signal $x_2 = \textrm{INV}(\mathbf{X}_2, \phase_x)$, where $\mathbf{X}_2 = \mathbf{X} - \sum_{l \in L_{c, x}} \mathbf{X}_l$. We define faithfulness of the interpretation to classifier $f$ for sample $x$ with:
\begin{equation}
    \textrm{FF}_x = f(x)_c - f(x_2)_c
\end{equation}
where $f(x)_c, f(x_2)_c$ denote the output probabilities for class $c$. It should be noted that this strategy to simulate removal may introduce artifacts in the input that can affect the classifier's output unpredictably. Also, interpretations on samples with poor fidelity can lead to negative $\textrm{FF}_x$. Both of these observations point to the potential instability and outlying values for this metric. Thus, we report the final faithfulness of the system as median of $\textrm{FF}_x$ over test set, denoted by $\textrm{FF}_{\textrm{median}}$. A positive $\textrm{FF}_{\textrm{median}}$ would signify that interpretations generally tend to be faithful to the classifier. 

As already discussed, it is not possible to measure faithfulness for concept-based \textit{post-hoc} interpretability approaches. While measurement for input attribution based approaches is possible, the interpretations themselves and the feature removal strategies are different, making comparisons with our system significantly less meaningful. We thus compare our faithfulness against a \textit{Random Baseline}, wherein the less-important components, those not present in $L_{c,x}$, are randomly removed. To compare fairly, we remove the same number of components that are present in $L_{c, x}$ on average. This would validate that, if the interpreter selects \textit{truly} important components for the classifier's decision, then randomly removing the less important ones should not cause a drop in the predicted class probability. 

We also emphasize at this point that works related to audio interpretability (see Sec. \ref{related:audio_intp}), are not suitable for comparison on these metrics. Particularly, APNet \cite{apnet} is not designed for \textit{post-hoc} interpretations. AudioLIME \cite{aud-lime-1} is not applicable on our tasks as it requires known predefined audio sources. Moreover, SLIME \cite{slime-2} and AudioLIME still rely on LIME \cite{lime} for interpretations. It is a feature-attribution method that approximates a classifier for \textit{each} sample separately. As discussed before, these characteristics are not suitable for comparison on our metrics.

Separate from the quantitative metrics, we conducted a subjective evaluation to evaluate quality and understandability of interpretations. Our design for the same was based on qualitative understanding of saliency maps for images. Attribution maps in images are qualitatively judged by observing the visual overlap in input with the given class being interpreted. In similar spirit, our design was based on providing the user with input and class being interpreted and asking them to rate auditory overlap of the interpretation and part of input audio corresponding to the class. Further details and results are covered in the next section. Apart from evaluating understandability, we also extensively analyze our interpretations qualitatively.

\noindent \textbf{Metrics and baselines (By-design)}. For by-design interpretation, the faithfulness metric is much less significant. This is because the final classification output is generated by the interpreter itself and thus faithfulness is ensured by-design. The classification performance of the interpreter is the primary metric, similar in spirit to fidelity evaluation for post-hoc interpretations. We compare this with several baselines to (i) benchmark performance of our by-design interpretable network, and (ii) to evaluate the two key modifications introduced in the learning problem while extending from post-hoc to by-design interpretation (section \ref{approach:by-design}). Specifically, the hidden layers of $f$ are not pre-trained on the given dataset in by-design problem and updated jointly with interpreter layers. And secondly, applying an additional classification loss on $f(x)$ to affect the hidden layers. The various baselines and the reasons to include them are the following:
\begin{itemize}
    \item Audio prototypical networks (APNet) \cite{apnet} act as a primary baseline from literature. It is an audio processing by-design interpretable network. While it generates interpretation differently from us, it is the only system in the literature addressing by-design interpretation for audio modality. Note that the dedicated post-hoc interpretation systems VIBI and SLIME are not relevant for this problem. For fair comparison, we use the same number of prototypes in their network as our number of components.  

    \item In order to ascertain that using a CNN based representation for NMF offer advantage over typical NMF based representations in terms of prediction performance, we also evaluate performance of two NMF variants: Unsupervised NMF based classification and the Task-driven Dictionary Learning (TDL)-NMF model \cite{bisot2017feature}. The unsupervised NMF model simply learns a dictionary on training data, computes average time activations on test samples and makes predictions using a linear model. The TDL-NMF model instead updates the initial learnt dictionary with classification loss from the linear model and thus learns them jointly. For both the systems, we experiment with use of two data types to learn NMF-dictionaries. The first is log-magnitude spectrograms and second is power mel-spectrogram (with a square root transformation). We vary dictionary sizes from 64 to 512 components and report results for best performance.  

    \item Given the framework level similarities between FLINT and L2I, we also evaluate the performance of the FLINT interpreter when trained for by-design interpretation. As before, we again emphasize that FLINT is not suitable for audio interpretations, but provides a interpretable network design for comparison of performance.  

    \item Variants of L2I: We denote our proposed version of L2I for by-design interpretation as $\textrm{L2I}_{\textrm{BD}}$ w/ $\Theta_{\textsc{att}}$. We further evaluate two variants of our proposed classification network $g(x)$. The first variant ``$\textrm{L2I}_{\textrm{BD}}$-NoPred" does not include a classification loss applied to $f(x)$ and instead applies it directly to $g(x)$. The second variant ``L2I-PostHoc" is simply the interpreter trained for post-hoc interpretation. We compare with these variants to gain perspective on effect of differences between our formulations of post-hoc and by-design problems.  
\end{itemize}
The implementation details of all the baselines (post-hoc and by-design) can be found on our companion website.\footnote{\url{https://jayneelparekh.github.io/listen2interpretV2/} \label{footnote 1}}


\section{Results and discussion}

\subsection{Post-hoc Interpretation}
\subsubsection{Fidelity}
As discussed previously, to quantify fidelity, we use the same respective metrics as done to benchmark classifier performance but evaluate them for interpreter output w.r.t classifier output. For ESC-50, mean and standard deviation of top-$k$ fidelity is calculated over the 5 folds. We show these results for $k=1, 5$.  
For SONYC-UST, we report the macro-AUPRC, micro-AUPRC and max-F1 for the interpreter output w.r.t classifier. For fairness, we ignore the class `non-machinery impact' from all class-wise evaluations involved in fidelity (\textit{i.e.} macro-AUPRC) or faithfulness. This is because the classifier predicts only one sample in test set with positive label for this class, causing AUPRC scores to vary widely for different interpreters. 
For OpenMIC-2018, we report the Fidelity weighted F1-score for each system. All the above results are available in Tab. \ref{fidelity_all}. 

Among the four systems, VIBI performs the worst in terms of fidelity. This is very likely because it treats the classifier as a black-box, while the other three systems access its hidden representations. This strongly indicates that accessing hidden layers can be beneficial for fidelity of interpreters. While on ESC50, FLINT achieves the best fidelity, L2I w/ $\Theta_{\text{ATT}}$ outperforms all systems on the other datasets. It should be noted that our system variants distinctly hold the advantage of generating listenable interpretations over FLINT and VIBI. Nevertheless, these systems form strong baselines for fidelity and the results demonstrate that our interpreter can generate high-fidelity \textit{post-hoc} interpretations. Moreover, its design is flexible w.r.t different pooling functions.

\begin{table*}
    \setlength{\tabcolsep}{7pt}
	\centering
    \resizebox{0.85\textwidth}{!}{%
	\begin{tabular}{l c c c c c} 
		\toprule
		& \multicolumn{2}{c}{ESC-50 (in \%)} & \multicolumn{2}{c}{SONYC-UST} & \multicolumn{1}{c}{OpenMIC-2018}\\
		\cmidrule[1pt](lr){2-3}
            \cmidrule[1pt](lr){4-5}
            \cmidrule[1pt](lr){6-6}
		System & top-1 & top-5 & macro-AUPRC & micro-AUPRC & avg-weighted-F1 \\ [0.5ex] 
		\midrule
		L2I w/ $\Theta_{\textsc{att}}$ & 65.7 $\pm$ 2.8 & 88.2 $\pm$ 1.7 & \bf{0.909 $\pm$ 0.011} & \bf{0.917 $\pm$ 0.008} & \bf{0.920 $\pm$ 0.004} \\ 
            L2I w/ $\Theta_{\textsc{max}}$ & 73.3 $\pm$ 2.3 & 92.7 $\pm$ 1.2 & 0.866 $\pm$ 0.014  & 0.913 $\pm$ 0.012 & 0.906 $\pm$ 0.004 \\
            \midrule
            FLINT \cite{flint} & \bf{73.5 $\pm$ 2.3} & \bf{93.4 $\pm$ 0.9} & 0.816 $\pm$ 0.013 & 0.907 $\pm$ 0.011 & 0.907 $\pm$ 0.004 \\
            VIBI \cite{vibi} & 27.7 $\pm$ 2.3 & 53.0 $\pm$ 1.8 & 0.608 $\pm$ 0.027  & 0.575 $\pm$ 0.019 & 0.581 $\pm$ 0.037 \\
 		\bottomrule
	\end{tabular}
     }
	\caption{Fidelity results for the interpreter w.r.t classifier's output on all datasets. We report top-1 and top-5 fidelity (in \%) for ESC-50 (all five folds), AUPRC-based metrics for SONYC-UST and weighted F1-score averaged over all classes for OpenMIC-2018. All results contain mean and variance over three runs. Values in bold indicate maximum of the metric among all the evaluated systems (incl. baselines).}
    \label{fidelity_all}
\end{table*}

\subsubsection{Faithfulness}
In Table \ref{faithfulness_esc50}, we report median faithfulness $\textrm{FF}_{\textrm{median}}$ on ESC-50 for our primary system L2I w/ $\Theta_{\textsc{ATT}}$ at different thresholds $\tau$ averaged over the five folds. Smaller $\tau$ corresponds to higher $|L_{c, x}|$, which denotes the number of components being used for generating interpretations. Thus, for Random Baseline, we report $\textrm{FF}_{\textrm{median}}$ at the lowest threshold $\tau=0.1$, to ensure removal of maximal number of components. To recall the definition of Random Baseline, please refer to Sec. \ref{metrics}. $\textrm{FF}_{\textrm{median}}$ for L2I w/ $\Theta_{\textsc{ATT}}$ is positive for all thresholds. It is also significantly higher than the Random Baseline, indicating faithfulness of interpretations.

\begin{figure}[!t]
\centering
\includegraphics[width=0.42\textwidth]{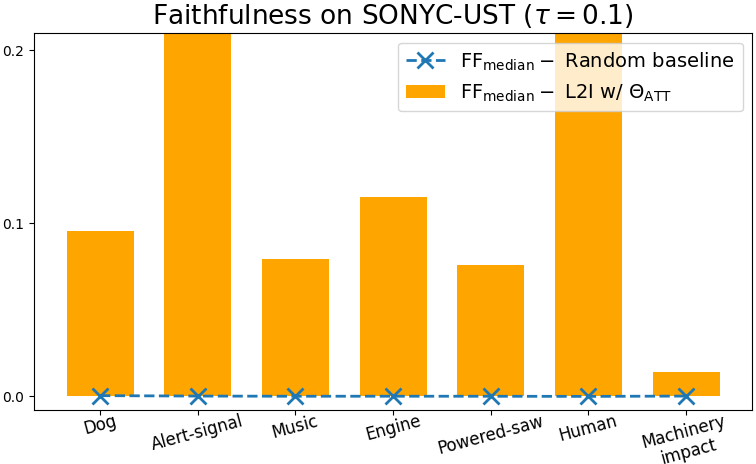}
\caption{Faithfulness (absolute drop in probability value) results for SONYC-UST arranged class-wise for threshold, $\tau=0.1$}
\label{fig_faithfulness_sony}
\end{figure}

\begin{figure}[!t]
\centering
\includegraphics[width=0.49\textwidth]{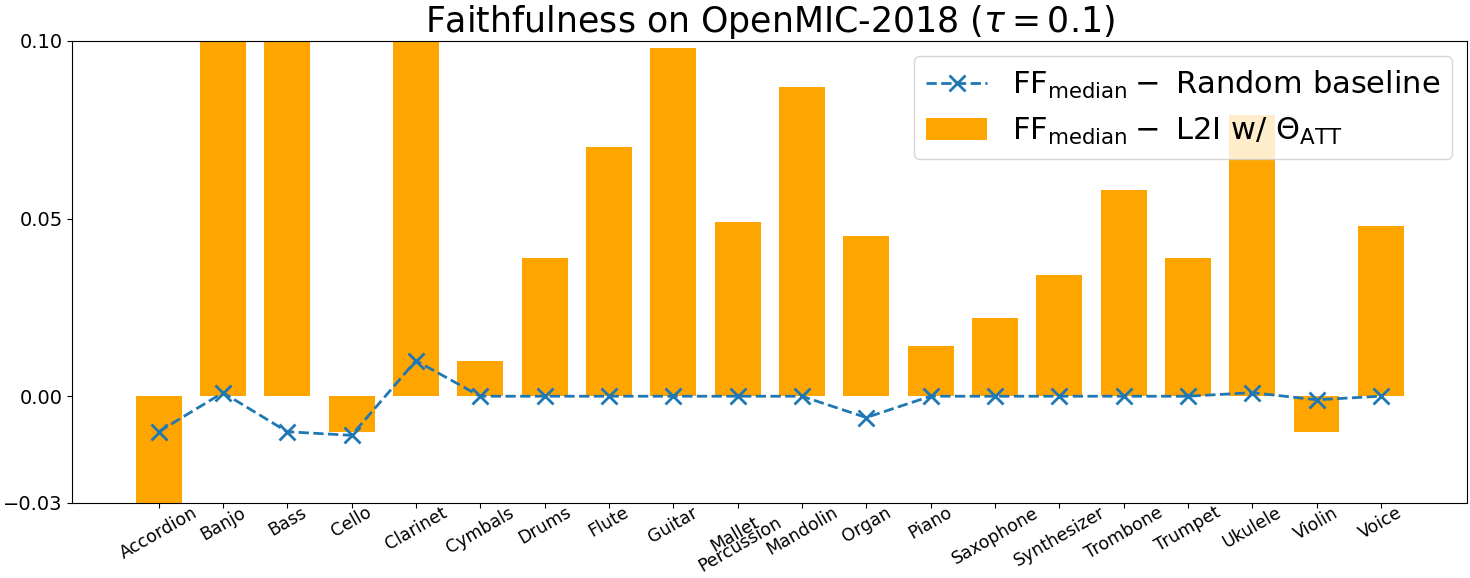}
\caption{Faithfulness (absolute drop in probability value) results for OpenMIC-2018 arranged class-wise for threshold, $\tau=0.1$}
\label{fig_faithfulness_omic}
\end{figure}

The results for class-wise faithfulness on SONYC-UST and OpenMIC are illustrated in Fig. \ref{fig_faithfulness_sony} and \ref{fig_faithfulness_omic} respectively. We show $\textrm{FF}_{\text{median}}$ (absolute drop in probability) for our system and the Random Baseline. For most classes, interpretations can be considered faithful, with a significantly positive median compared to random baseline results, which are very close to 0.

\begin{table}[!t]
\setlength{\tabcolsep}{7pt}
	\centering
	\resizebox{0.4\textwidth}{!}{%
	\begin{tabular}[t]{l c c} 
		\toprule
		System & Threshold $\tau$ & $\text{FF}_\text{median}$ \\ [0.5ex] 
		\midrule
		\multirow{ 5}{*}{L2I w/ $\Theta_{\textsc{att}}$} & $\tau=0.9$ &  0.002 \\
		 & $\tau=0.7$ &  0.004 \\
		 & $\tau=0.5$ & 0.012\\
		 & $\tau=0.3$ &  0.040\\
		 & $\tau=0.1$ &  0.113\\
		\midrule
		Random Baseline & $\tau=0.1$ & $<10^{-4}$ \\
		\bottomrule\\
	\end{tabular}}
	\caption{Faithfulness results on ESC-50 for different thresholds, $\tau$. We report $\text{FF}_\text{median}$ for proposed L2I w/ $\Theta_{\textsc{att}}$ and the Random Baseline.}
    \label{faithfulness_esc50}
\end{table}

\subsubsection{Subjective evaluation} 
\begin{figure}[!t]
\centering
\includegraphics[width=0.48\textwidth]{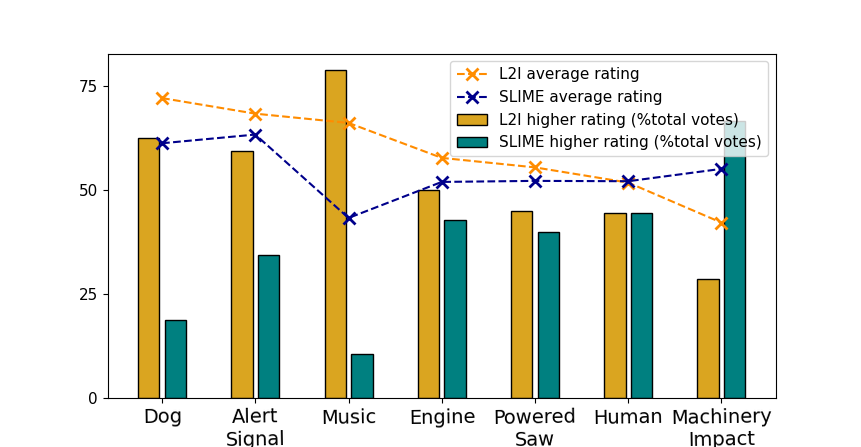}
\caption{Subjective evaluation results. Average scores for L2I and SLIME and fraction of votes in favour of each system}
\label{fig_sub_eval}
\end{figure}

The test was conducted with 15 participants. Each participant was provided with 10 input samples, a predicted class by the classifier for each sample and the corresponding interpretation audios from SLIME and L2I. They were asked to rate the interpretations on a scale of 0-100 for the following question: ``\textit{How well does the interpretation correspond to the part of input audio associated with the given class?}". The 10 samples were randomly selected from a set of 36 (5-6 random test examples per class). For each sample, we ensured that the predicted class was both, present in the ground-truth and audible in input. Class-wise preference results and average ratings are shown in Fig. \ref{fig_sub_eval}. L2I is preferred for 'music', 'dog' \& 'alert-signal', SLIME is preferred for 'machinery-impact', no clear preference for others.

\subsection{Qualitative analysis of interpretations}
Qualitatively we observe that our interpretations are capable of emphasizing the object of interest and are insightful for an end-user to understand the classifier's prediction. We share multiple examples on our companion website.\textsuperscript{\ref{footnote 1}} Samples in case of SONYC-UST and OpenMIC are often already challenging with the presence of other sources of audio. In case of ESC50, to create more interesting and challenging scenarios we devise an experiment described below

\noindent \textbf{Audio corruption experiment: interpretability illustration}.~  For ESC50, we generate interpretations after corrupting the testing data for fold--1 in two different ways (i) either with white noise at 0dB SNR (signal-to-noise ratio), (ii) or mixing it with a sample of a different class. It should be noted that in both these cases the system is exactly the same as before and \textbf{not} trained with corrupted samples. Some examples, covering both types of corruptions are shared on our companion website.\textsuperscript{\ref{footnote 1}}. 

For SONYC-UST, we observe good interpretations for classes `alert-signal', `dog' and `music'. For them, the background noise is significantly suppressed and the interpretations mainly focus on the object of interest. Interpretations for class `human' are also able to suppress noise to a certain extent and focus on parts of human voices. However, for this class, we found presence of some signal from other audio sources too. 
For the remaining classes, namely `Engine', `Powered-saw' and 'Machinery-impact' the quality of the interpretation is more sample dependent. This is due to their acoustic similarity with the background noise. We provide example interpretations for SONYC-UST on our companion website.\textsuperscript{\ref{footnote 1}} 

The third dataset OpenMIC-2018, offers challenges under unique scenarios. Unlike SONYC-UST while it does not face issue of noise in data, it faces the hurdle of a strong overlap between instruments. This is because their onsets are often aligned by beats of the musical piece. This increases difficulty of filtering the signal of interest. Even with the greater complexity, the interpretations in many cases are able to emphasize the class of interest. Classes with relatively unique sounds such as `Bass' or `Mallet-percussion' are very well extracted. String like instruments including Violin and Guitar are also generally emphasized well.    

\begin{figure}[!ht]
    \centering
    \includegraphics[width=0.85\columnwidth]{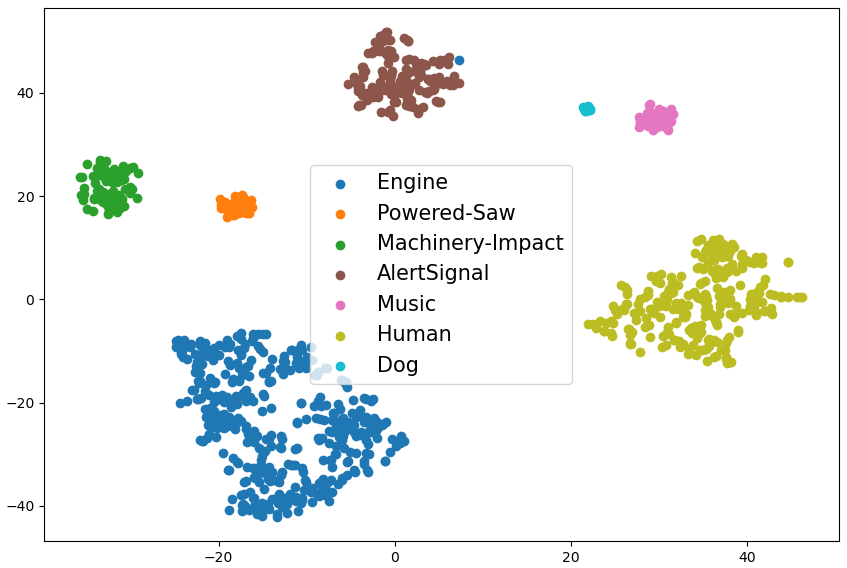}
    \includegraphics[width=0.98\columnwidth]{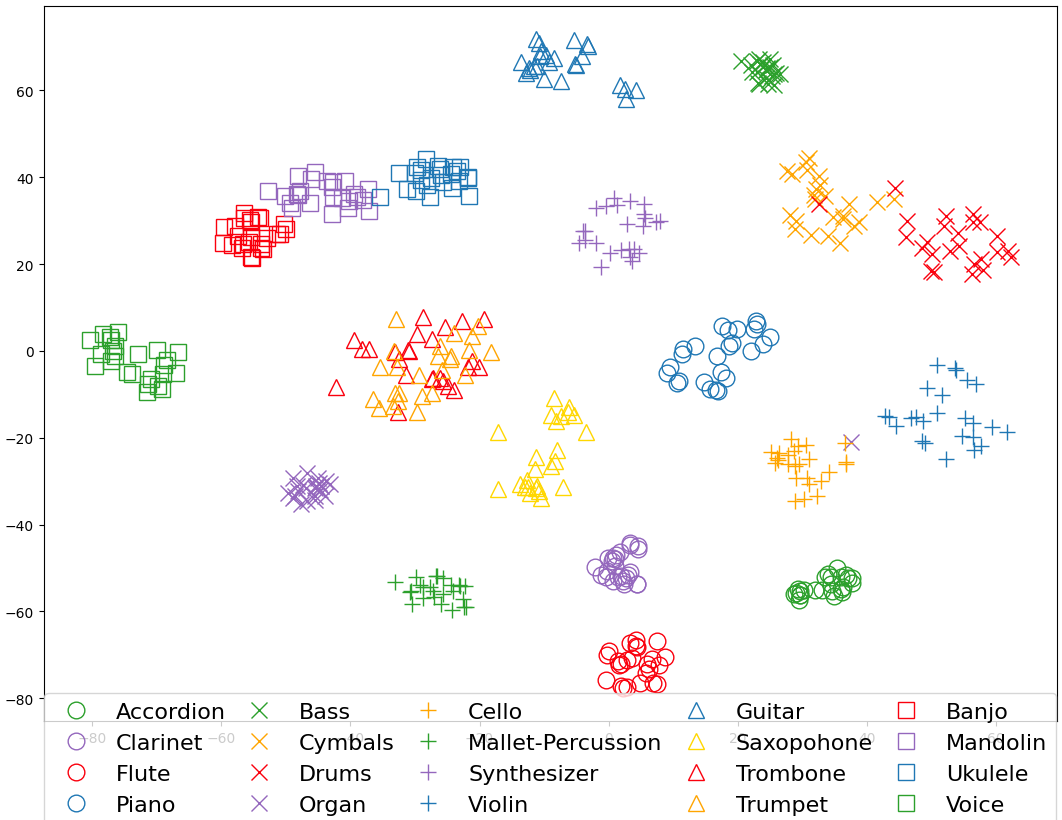}
    \caption{Visualized relevances (following a t-SNE transformation) of generated interpretations on test sets of SONYC-UST (top) and OpenMIC-2018 (bottom), colour-coded according to interpreted class. For clarity in case of OpenMIC, we show up to random 25 interpretations of a class.}
    \label{tsne_rel}
\end{figure}

\noindent \textbf{Coherence of interpretations}.~ We visualize interpretations generated on the test set for SONYC-UST and OpenMIC-2018 by clustering relevance vectors. Specifically, we compute the vector $r_{c, x} \in \mathbb{R}^K$ which contains relevances of all components in prediction for class $c$ for sample $x$. The relevance vectors are collected for each test sample $x$ and its predicted class $c$. We then apply a t-SNE \cite{tsne} transformation to 2D for visualization. This is shown in Fig. \ref{tsne_rel}. Each point is labeled/colored according to the class for which we generate the interpretation. Interpretations for any single class are coherent and similar to each other. This is to some extent a positive consequence of global weight matrix in $\Theta$. Moreover, globally it can be observed that classes like 'Machinery-impact' and 'Powered-Saw' have similar relevances which are to some extent close to 'Engine'. This is to be expected as these classes are acoustically similar. 'Dog' and 'Music' are also close in this space, likely due to the often periodic nature of barks or beats. The visualization for OpenMIC is arguably even more interesting because of larger number of classes and several inter-class relationships. Various sets of similar instruments end-up as clusters in proximity of each other. The examples include `Cello-Violin', `Drums-Cymbals', `Clarinet-Flute', `Ukulele-Mandolin-Banjo', `Trombone-Trumpet-Saxophone'. Moreover, the meaningfulness of clustering also extends to higher-level of grouping. For example, the data is partitioned so as the string-based, wind-based, or percussion instruments are close to each other within their respective groups. This indicates that the interpreter's representations of what constitutes sound of an instrument aligns to some extent to human understanding. 

\subsection{Ablation studies} 

Tab. \ref{ablation_hidden_layer} and Tab. \ref{ablation_loss} present ablation studies for loss hyperparameters and choice of hidden layers. The values in bold indicate our current choices for post-hoc interpretation. The metrics and loss values given here are for a single run.

\begin{table}[!t]
\setlength{\tabcolsep}{8pt}
	\centering
	\resizebox{0.36\textwidth}{!}{%
	\begin{tabular}{l c c c} 
		\toprule
		ConvBlocks & $\bmL_{\textrm{NMF}}$ & $\bmL_{\textrm{FID}}$ & top-1 \\ [0.5ex] 
		\midrule
          B3 & 0.104 & 1.788 & 53.0 \\
          B6 & 0.118 & 1.698 & 57.8\\
          B2+B3 & 0.093 & 1.966 & 51.8 \\
		 B5+B6 & 0.103 & 1.572 & 61.5 \\
          \textbf{B4+B5+B6} & \textbf{0.079} & \textbf{1.546} & \textbf{65.5} \\
          \midrule
		 Input & 0.102 & 2.384 & 34.5\\
		\bottomrule\\
	\end{tabular}}
	\caption{Ablation study for hidden layers: loss values on ESC50 (fold 1) test set for different subsets of hidden layers. Current choice indicated in bold.}
    \label{ablation_hidden_layer}
\end{table}

\begin{table}[!t]
\setlength{\tabcolsep}{7pt}
	\centering
	\resizebox{0.44\textwidth}{!}{%
	\begin{tabular}{c c c c c} 
		\toprule
		$\alpha$ & $\beta$ & $\bmL_{\textrm{NMF}}$ & $\bmL_{\textrm{FID}}$ & macro-AUPRC \\ [0.5ex] 
		\midrule
		 \textbf{10.0} & \textbf{0.8} & \textbf{0.028} & \textbf{0.386} & \textbf{0.900}\\
		 10.0 & 8.0 & 0.048 & 0.386 & 0.879\\
         10.0 & 0.08 & 0.028 & 0.388 & 0.876\\
         1.0 & 0.8 & 0.045 & 0.375 & 0.921\\
         100.0 & 0.8 & 0.027 & 0.445 & 0.612\\
		\bottomrule\\
	\end{tabular}}
	\caption{Ablation study for loss hyperparameters: loss values on SONCY-UST test set for different weights of loss functions. Current choice indicated in bold.}
    \label{ablation_loss}
\end{table}

\noindent Selecting the \textbf{hidden layers} of the classifier that should be accessed by the interpreter is an important choice. At first glance, this model selection task might appear to be computationally too expensive as total possible choices is exponential in number of hidden layers. However, practical considerations can heavily reduce the search space. An upper bound to the number of layers could be set depending upon the desired size of interpreter. In our experiments throughout the paper, we limited ourselves to at most 3 layers. Crucially, layers close to the output are more favourable, for multiple reasons. They generally result in better fidelity and inherently tie the interpreter much closer to the output of classifier. Moreover, the latter layers are also expected to capture higher level features. We illustrate how selecting different subsets of hidden layers affects optimization of our fidelity and reconstruction loss by doing an ablation study. It's results are reported in table \ref{ablation_hidden_layer}. The classifier consists of 6 major convolutional blocks (B1--B6).

\noindent \textbf{Loss weights}. We illustrate the effect of varying loss weights on optimization in table \ref{ablation_loss}. Too high emphasis on $\bmL_{\textrm{NMF}}$, that is, high $\alpha$ can hurt the fidelity of interpreter while a high $\beta$ (sparsity loss) can result in poorer reconstruction. Importantly, there is a good range of values wherein the system can be regarded as operating reasonably. 

\begin{figure}
    \centering
    \includegraphics[width=0.37\columnwidth]{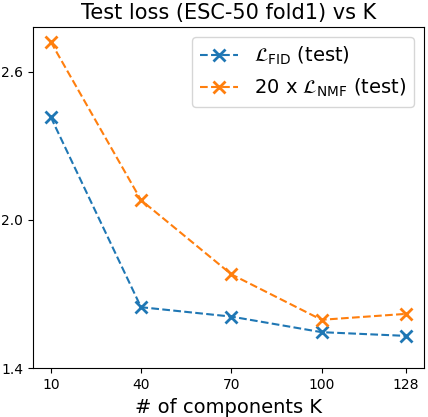}
    ~
    \includegraphics[width=0.591\columnwidth]{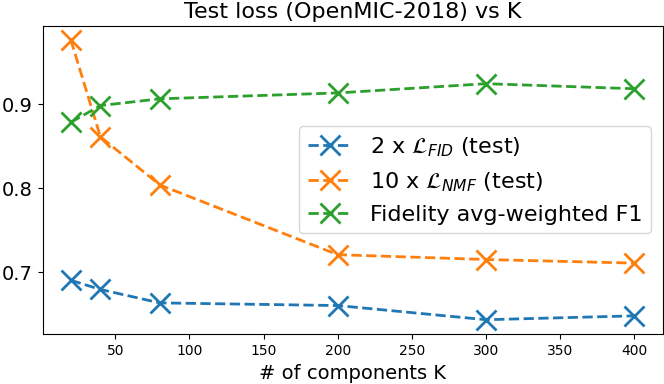}
    \caption{Ablation study for number of components. Loss values on test data for ESC-50 and OpenMIC-2018.}
    \label{fig_loss_vs_K}
\end{figure}

\noindent \textbf{Number of components}. Choosing $K$, also known as order estimation, is typically data and application dependent. It controls the granularity of the discovered audio spectral patterns. 
Determining the optimal value has been a long standing problem within the NMF community \cite{tan2012automatic}. Our choice for this parameter was guided by three main factors: 
\begin{itemize}
    \item Choices made previously in literature for similar pre-learning of $\mathbf{W}$ \cite{bisot2017feature}, who demonstrated reasonable acoustic scene classification results with a dictionary size of $K=128$. We used this as a reference to guide our choice.
    \item Dataset specific details which include number of classes, samples for each class, variability of recordings etc. For eg. acoustic variability of ESC-50 (larger number of classes), prompted us to use a dictionary of larger size compared to SONYC-UST. We use highest number of components for OpenMIC, which has largest dataset size among the three and reasonably high acoustic variability.  
    \item When tracking loss values for different $K$, we observed a plateauing effect for larger dictionary sizes as illustrated in Fig. \ref{fig_loss_vs_K} for ESC-50 and OpenMIC-2018. In case of OpenMIC, this effect is prominent for reconstruction loss $\bmL_{\text{NMF}}$. The fidelity remains high even for small $K$ . 
\end{itemize}

\subsection{By-design Interpretation}
\label{experiments:by-design}

The performance of all systems is given in Tab. \ref{performance_all}. We compute the same metrics as used to evaluate the classifiers for each dataset. Mean performance along with variance across 3 runs is reported. We make the following key observations:
\begin{itemize}
    \item Among the \textbf{interpretable neural networks for audio}, $\textrm{L2I}_{\textrm{BD}}$ w/ $\Theta_{\textsc{att}}$ clearly outperforms APNet. The size of the models plays an important role in this. L2I learns with the help of a network architecture that feeds it with higher quality representations for prediction compared to architecture in APNet. It is generally able to sustain a comparable performance w.r.t the base network BASE-$f$ while imposing an interpretable structure for final prediction model.

    \item \textbf{Comparison with NMF baselines}. While TDL-NMF performs better than unsupervised-NMF, L2I variants are noticeably better than both. This highlights a unique advantage of combining NMF representations with deep neural network representations, wherein, the NMF structure leads to interpretability and using deep networks as source provides higher prediction performance compared to directly using NMF activations generated from input.   

    \item We also validate our design of training procedure for by-design interpretable network $g(x)$, by comparing it with the two variants of proposed system, L2I-PostHoc and $\textrm{L2I}_{\textrm{BD}}$-NoPred. The performance of $\textrm{L2I}_{\textrm{BD}}$ w/ $\Theta_{\textsc{att}}$ compared to L2I-Posthoc highlights that $g(x)$ tends to perform better when hidden layers of $f$ are trained jointly with interpreter. $\textrm{L2I}_{\textrm{BD}}$-NoPred performs the worst among the three, emphasizing the benefits of updating the hidden layers of $f$ with classification loss imposed on $f(x)$ rather than on $g(x)$.  
\end{itemize}

\begin{table*}
    \setlength{\tabcolsep}{7.5pt}
	\centering
    \resizebox{0.7\textwidth}{!}{%
	\begin{tabular}{l c c c} 
		\toprule
		& \multicolumn{1}{c}{ESC-50 (in \%)} & \multicolumn{1}{c}{SONYC-UST} & \multicolumn{1}{c}{OpenMIC-2018}\\
		\cmidrule[1pt](lr){2-2}
            \cmidrule[1pt](lr){3-3}
            \cmidrule[1pt](lr){4-4}
		System & accuracy & macro-AUPRC & avg-weighted-F1 \\ [0.5ex] 
		\midrule
		$\textrm{L2I}_{\textrm{BD}}$ w/ $\Theta_{\textsc{att}}$ & \textbf{70.1 $\pm$ 1.5} & \textbf{0.581 + 0.008} & \bf{0.825 $\pm$ 0.005}\\ 
            APNet \cite{apnet} & 63.6 $\pm$ 1.7  & 0.422 $\pm$ 0.012 & 0.563 $\pm$ 0.025 \\
            Unsupervised-NMF & 39.4 $\pm$ 2.3 & 0.373 $\pm$ 0.006 & 0.659 $\pm$ 0.018\\
            TD-NMF \cite{bisot2017feature} & 46.7 $\pm$ 2.7 & 0.431 $\pm$ 0.018 & 0.699 $\pm$ 0.012\\
            L2I--Posthoc & 65.4 $\pm$ 3.4 & 0.567 $\pm$ 0.007 & \textbf{0.825 $\pm$ 0.003}\\
            $\textrm{L2I}_{\textrm{BD}}$--NoPred & 64.4 $\pm$ 1.1 & 0.563 $\pm$ 0.004 & 0.746 $\pm$ 0.006\\
            \midrule
            FLINT \cite{flint} & 75.3 $\pm$ 3.6 & 0.556 $\pm$ 0.008 & 0.827 $\pm$ 0.002 \\
            BASE-$f$ & 82.5 & 0.601 & 0.831\\
 		\bottomrule
	\end{tabular}
     }
	\caption{Classification performance for by-design interpretation. The evaluated systems include our proposed by-design interpretable network, denoted as $\textrm{L2I}_{\textrm{BD}}$ w/ $\Theta_{\textsc{att}}$, its variant with modified loss function ($\textrm{L2I}_{\textrm{BD}}$--NoPred), interpreter trained for post-hoc interpretation (L2I--Posthoc), classification models based on traditional NMF (unsupervised NMF and TD-NMF) and audio prototypical network APNet. The base classification network used for post-hoc interpretation (BASE-$f$) and FLINT are used as references for high performance networks not suitable for audio interpretation.}
    \label{performance_all}
\end{table*}

\section{Conclusion}


We have presented a framework to tackle both post-hoc and by-design for audio classification networks. To this end, a novel interpreter is designed with the key idea of using an NMF-inspired regularizer. This enables listenable concept-based interpretations. We motivate listenability as an important attribute for audio interpretability. Efficacy of the proposed framework is established through extensive qualitative and quantitative experimentation. In particular, we quantitatively evaluate both post-hoc and by-design interpretations on three popular datasets pertaining to audio event and music instrument recognition tasks. We perform a user-study to confirm usefulness of our interpretations. In addition, through a visualization of the generated interpretations, we show that they are coherent across samples from different classes and cluster in a fashion that aligns well with human understanding of sound. Further works concern the extension of this framework to other machine learning audio-based tasks.

\bibliography{references_short}
\bibliographystyle{IEEEtran}

\end{document}